**Electronics-free passive ultrasonic communication link for deep-tissue sensor implants**

**Authors**

Umut Can Yener[1], Alp Timucin Toymus[1], Kivanc Esat[2], Mehdi Alem[2], Levent Beker*[1,3]

**Affiliations**

[1]Department of Mechanical Engineering, Koç University, Rumelifeneri Yolu, Sarıyer, Istanbul, 34450, Turkey

[2]Zurich Instruments, Ltd., Technoparkstr. 1, 8005 Zurich, Switzerland.

[3]Department of Biomedical Sciences and Engineering, Koç University, Rumelifeneri Yolu, Sarıyer, Istanbul, 34450, Turkey

*Corresponding author: Levent Beker – lbeker@ku.edu.tr

**Abstract**

Wireless communication is critical for the non-invasive in-situ monitoring of vital signs in deep tissues. Wireless ultrasonic links demonstrated so far solved the shortcomings of electromagnetic wave-based communication methods. However, they either require highly customized and complex designs for the implant electronics or rely on physical changes in implanted metamaterials. Here, we report a wireless, passive, frequency-based, and electronics-free ultrasonic communication method for deep tissue sensor implants. The device consists of a piezoelectric crystal used as the ultrasonic antenna, considerably simplifying the implant design, and can be integrated with any capacitive implantable sensor. We demonstrate the applicability of the passive communication method using a commercial pressure sensor at a depth of 5 cm with decent precision and accuracy.

**Introduction**

Monitoring vital signs in deep tissues is a major challenge. Many diseases and disorders occur deep within the tissue, which limits the effectiveness of the measurement techniques that rely on superficial signs[1–3]. Medical imaging techniques such as ultrasonography and magnetic resonance imaging are helpful in getting clinically relevant information from deep tissues non-invasively but they are not suitable for cases where continuous in-situ monitoring is necessary[4,5]. For such cases, it is possible to monitor signs deep within the body using invasive instruments like catheters[6–8]. Nonetheless, these alternatives are restricted to clinical settings, require trained personnel, and are costly for the healthcare system. Moreover, the invasive nature of these techniques holds the risk of causing severe discomfort, bleeding, scarring, and bacterial infections; thus preventing frequent and continuous monitoring[9–13]. Therefore, monitoring vital signs deep within the body via wireless implants is an attractive solution.

As an alternative to the clinically available invasive measurement techniques, medical implants have demonstrated the ability to sense signals subcutaneously. Unlike early examples that use percutaneous wired connections to external devices, recent implantable devices use wireless power and data-transferring technologies such as inductive coupling[14–17], magnetic resonance coupling[18–20], near-field communication[21–23], and Bluetooth[24–26]. However, high tissue attenuation and low penetration depth associated with electromagnetic waves either limit the implantation depth for these devices or necessitate wired connections between the deeply implanted sensor and subcutaneously implanted antenna (Supplementary Fig. 1), which introduce similar risks associated with wired connections[27,28].

Power and data transferring via ultrasound (US) have recently attracted significant attention for deep-tissue implantable devices. US offers millimeter (below 1.5 MHz) and submillimeter (above 1.5 MHz) wavelengths at the diagnostic US frequency range, which achieves better coupling with miniaturized implants when compared to electromagnetic (EM) waves (wavelength of 25 mm at 2 GHz). Additionally, US offers lower attenuation in soft tissues (~1.5 dBcm$^{-1}$ at 2 MHz for US and ~10 dBcm$^{-1}$ at 2 GHz for EM) which results in higher penetration depth and less tissue heating[29]. For these reasons, several studies have demonstrated implantable devices with US power and data links[30–38] for various applications including wireless recording of the nervous system, and real-time tissue oxygenation and temperature monitoring. However, all devices with US data links demonstrated thus far need electronic components for power harvesting and active digital communication. As a result, these devices consist of capacitor elements to store energy, rectifier circuits to harvest power, and custom integrated circuit (IC) chips to modulate sensor data to the backscattered ultrasound waves and to switch between powering and data transferring modes, which lead to heavily customized and complex implant electronics design and fabrication. Alternatively, several studies have demonstrated electronics-free ultrasonic communication using implanted metamaterials[39–42]. However, these devices cannot be used for general-purpose communication methods as they rely on application-specific deformations or property changes on the metamaterials. Furthermore, a recent study has demonstrated a passive ultrasound pressure sensing device leveraging the amplitude changes in impedance[43]. However, amplitude-based approaches have limited applicability as they are highly susceptible to disturbances in the acoustic coupling. Therefore, a frequency-based alternative would be an important step towards reliable passive ultrasonic communication.

Here we present an elegantly simple frequency-based passive ultrasonic communication (PUC) method, to enable electronics-free data transfer from deep tissue capacitive sensor implants for real-time monitoring applications. The implantable device consists of an ultrasonic antenna comprised of a single piezoceramic crystal and an implantable capacitive sensor. The PUC method leverages the changes in the resonance characteristics of the ultrasonic antenna due to the sensor's capacitance, which can be captured by an external interrogator ultrasonic transducer, thereby eliminating the necessity for custom IC chips and power harvesting circuits and considerably simplifying the implant design. The electronics-free communication system is demonstrated in-vitro with a commercial pressure sensor and is capable of detecting pressures in a clinically relevant range. The system presents a promising ultrasonic communication method for deep tissue applications that can be integrated with any implantable capacitive sensor.

## Results

**Device design and working principle.** The operating principle of the wireless and passive ultrasonic communication method is shown in Fig. 1a. The passive link is achieved between an external interrogator transducer and a piezoceramic US antenna connected to a capacitive sensor.

The interrogator transducer transmits ultrasound waves, which travel through the soft tissue and get reflected by the implantable US antenna along with the surrounding soft-tissue boundaries. The backscattered echo of the incident pulse is then captured by the interrogator transducer and converted to electrical signals to extract the sensor readings wirelessly.

An electrical equivalent circuit can be constructed for the US antenna and capacitive sensor, which comprises a van-Dyke Butterworth PZT model and a single capacitor, respectively (Fig. 1b). Monitored signs cause capacitive changes ($C_L$) in the implantable sensor. These changes directly affect the resonance characteristics of the whole device[44] (Supplementary Note 1). In particular, the anti-resonance frequency of the system shifts in correlation with the changes in the capacitive load that is connected to the US antenna (Fig. 1c, Supplementary Fig. 2 and 3). At this frequency, the incoming wave backscatters less resulting in reduced signal amplitude captured by the interrogator transducer. This effect is determined by the acoustic reflection coefficient as described in Supplementary Note 2. Therefore, exciting the US antenna by an external transducer at frequencies near the anticipated anti-resonance frequency, and recording the backscattered ultrasound waves result in a spectrum where the minimum point reveals the anti-resonance frequency of the implanted system, which can be used to find the capacitive load, and consecutively the deep tissue measurements performed by the implantable capacitive sensor (Fig. 1d).

The exploded view of the fabricated ultrasonic antenna is presented in Fig. 1e. The device consists of a millimeter-scale piezoceramic, a 200-µm thick flexible electronic board substrate that facilitates the electrical connections, and a biocompatible polymer coating for chemical and electrical isolation. Moreover, an air-backing design approach is adopted by creating an air cavity at the back of the PZT crystal to reduce the mechanical damping which helps the PZT crystal vibrate more freely. Additionally, the capacitive pressure sensor is shown soldered to the fabricated antenna device. Fig. 1f shows the fabrication flow of the device. First, the thin conductive substrate is patterned and cut, then the PZT crystal is soldered on the substrate and finally, the device is encapsulated for electrical and chemical isolation.

Considering the wavelength for efficient coupling to mm-scale US antenna and frequency-dependent attenuation of the ultrasound in soft tissues[45], the aspect ratio and thickness of the PZT crystal were determined such that the resonance and anti-resonance frequencies of the thickness mode operation are around 2 MHz. As a result, the US antenna device has a 5×7 mm$^2$ footprint and a total volume of 16 mm$^3$ (Fig. 1g).

**Electrical and mechanical device characterization of the device.** Fig. 2a shows a detailed Leach electroacoustic equivalent circuit model (ECM) of the ultrasonic antenna where the PZT crystal, backing layer, and acoustic medium are modeled as lossy transmission lines. The Leach model was selected due to its SPICE-friendly nature, unlike KLM, Mason, and Redwood circuit models which consist of either frequency-dependent transformers or negative valued capacitive elements[46–48].

The electrical impedance measurement of the device was compared to the Leach model simulation results in Fig. 2b. The measurement shows an anti-resonance frequency of 2.3 MHz as expected. Although the simulation results match the measurement at anti-resonance frequency, resonance profiles do not match due to the thickness resonance of the PZT crystal being unisolated from other vibration modes due to using unconventional PZT dimensions[49]. Nevertheless, the model-measurement mismatch is not crucial as the key is the precise detection of small changes in the capacitive load utilizing the anti-resonance frequency shifts. The measured and simulated effects of capacitive load on the impedance spectrum are shown in Fig. 2c,d. Here, the thickness mode anti-resonance frequency decreases as the load capacitance connected to the US antenna increases. The observed behavior supports the proposed method of passive communication.

The measured and simulated anti-resonance frequency shifts were compared in Fig. 2e, showing decent agreement. The ultrasonic antenna was further characterized using an acoustic frequency sweep test, presented in Fig. 2f, showing maximum acoustic pressure at 2.02 MHz, measured by a needle hydrophone 3 cm away from the antenna surface, which indicates that the resonance observed around 2 MHz is the thickness vibration mode of the antenna. Furthermore, the acoustic pressure field of the US antenna was measured using a motorized ultrasonic measurement system (Fig. 2g), revealing a -3dB diameter of 9.4 mm, 3 cm away from the source.

**Communication link characterization.** Fig. 3a shows the schematic block diagram of the communication hardware system. A duplexer is used to connect the interrogator transducer to the input and output channels of a lock-in amplifier, enabling transmit and receive operations from the same interrogator transducer, and hence facilitating pulse-echo measurements. The internal oscillator of the lock-in amplifier determines the frequency of the generated excitation signals which are then amplified by an RF amplifier. To demodulate the echo signal, the lock-in amplifier first mixes it with a sinusoidal waveform at the same frequency as the excitation signal and then applies the outcome to a low-pass filter which forms the envelope of the unprocessed time-domain signal. An unprocessed pulse-echo waveform and its demodulated form are shown in Fig. 3b. After the signal is demodulated, the lock-in amplifier obtains the average signal level within the demodulator trigger interval. These pulse-echo measurements are repeated for each frequency until the entire frequency interval is swept. The minimum point in the resulting frequency-dependent amplitude spectrum reveals the minimum acoustic reflection point, which is the anti-resonance frequency of the US antenna.

To confirm the effect of the presence of the US antenna, sweep operations were performed with and without the US antenna in front of the interrogator transducer. The resulting sweeper spectrum shows a clear valley when the US antenna is in front of the commercial interrogator transducer. In contrast, when a thin metal object is placed instead of the US antenna, the previously observed valley disappears (Fig. 3c). Moreover, the effect of the excitation signal duration and amplitude were analyzed. The results depicted in Fig. 3d showed that the valley of the sweep profile gets more pronounced as the signal duration increases. On the other hand, the amplitude of the excitation signal has no noticeable effect on the performance (Supplementary Fig. 5).

The PUC method was further analyzed for varying interrogator-antenna distance as shown in Fig. 3e. The distance only slightly affects the communication method results and it can be entirely compensated since the distance information can be extracted from the time of flight of the pulse-echo measurements. Moreover, the effect of resistive loads on the communication sweep results was analyzed in Fig. 4f-g and Supplementary Fig. 4. The results suggest that capacitive load is mainly responsible for the shifts, and the resistive effects work in conjunction as described in Supplementary Note 2.

Fig. 3h and i show the PUC sweep results for various capacitive loads connected to the US antenna in parallel. As the load capacitance increases from 0 to 120 pF, the valley in the backscattered ultrasound wave spectrum experiences a total shift of 18.56 kHz. The average standard deviation for measurements was calculated as 1.71 kHz. A $2^{nd}$ order polynomial curve fit reveals a sensitivity of 236 Hz/pF and 62 Hz/pF around 0 and 120 pF, respectively. Fig. 3j shows a violin plot representing the distribution of the measured anti-resonance frequencies for three capacitive load values, showing decent precision.

**In-vitro demonstration of the PUC system.** In order to demonstrate the applicability of the PUC system, a commercial sensor was integrated with the US antenna, which was then placed in a custom-designed pressure tank. Fig. 4a shows the schematic of the in-vitro experimental setup and Supplementary Fig. 6 shows the optical image of the setup. The air inlet and outlet ports of the tank were controlled manually to achieve the desired air pressure, while a commercial air pressure sensor was used to monitor the air pressure inside. The tank was partially filled with distilled water in order to facilitate the ultrasound wave propagation between the interrogator transducer and the US antenna, which were 5 cm apart from each other.

The capacitance of the integrated commercial sensor (FlexiForce A201, Tekscan) was measured under varying pressure levels (Fig. 4b), showing a sensitivity of approximately 5.83 pF/kPa at pressures below 20 kPa. Using the integrated system consisting of US antenna and sensor, the air pressure inside the tank was monitored using the electronics-free PUC method, and was compared to the commercial pressure sensor (NeuLog NUL-210) readings. Fig. 4c shows the communication sweeps performed at four different pressure levels before any post-processing steps, showing decent separation and precision. Moreover, Fig. 4d shows the mean and standard deviation of the anti-resonance frequencies extracted from the communication sweeps. Additionally, cyclic air pressure measurement results presented in Fig. 4e show the transient performance of the PUC method. During the transient experiment, the proposed system utilized a sweep resolution of 500 Hz resulting in a sample rate of approximately 1 Hz, limited by the hardware as discussed in the Methods section. Overall, the PUC method results demonstrate decent agreement with the pressure measured by the commercial sensor.

**Discussion**

In this work, we presented a simple, wireless, and passive frequency-based ultrasonic communication method for deep tissue sensor implants. The method utilizes ultrasound waves and leverages the changes in the resonance characteristics of the implanted US antenna in order to facilitate communication at greater implantation depths. The system consists of a single piezoceramic US antenna connected to a capacitive sensor in parallel. We demonstrated the performance of the method in vitro at a depth of 5 cm for medically relevant pressures[50–53] by integrating a commercial capacitive pressure sensor.

Compared to existing technologies that use ultrasound for establishing wireless communication links[30–43], the PUC method offers several advantages. First, the proposed method features only a single piezoceramic element, used as an ultrasonic antenna, soldered on a flexible conductive substrate. This enables simpler fabrication without needing any microfabrication processes. Second, the PUC system completely eliminates the need for custom-designed integrated circuits and other power-harvesting electronic components, which simplifies the implant design significantly. Finally, the PUC method offers flexibility in terms of sensor integration since any sensor with a capacitive sensing behavior can be used in conjunction with the US antenna.

We observed that the sweep results are affected by the angular and transverse misalignments between the antenna and the interrogator transducer. Therefore, future work will focus on analyzing the misalignment effects and formulating a generalized calibration algorithm that can diminish the misalignment susceptibility. In addition, the US antenna is fabricated from PZT, a material that contains lead. Although biocompatibility is critical for implantable systems, these concerns were not prioritized for this proof-of-concept work. However, lead-free alternatives like KNN[54] can be used to fabricate the US antenna as a future work. Furthermore, the antenna can be miniaturized using microfabrication processes to minimize tissue damage risks during implantation, and the sensor and antenna can be fabricated on the same substrate to achieve a monolithic implantable device. On the data processing side, the delay between the excitation and echo becomes only a few tens of microseconds, which limits the excitation duration to prevent overlap. Capturing the echo's envelope that expends over such short durations necessitates fast data conversion, demodulation, data transfer, and frequency determination. We anticipate that increasing the speed of these communication steps could lead to better signal quality, boosting the sensing precision and increasing the measurement speed beyond 2 Hz per frequency sweep. Alternative frequency sensing methods such as band-excitation using frequency chirps and pulse excitations can increase the frequency sweep rate as high as 1 kHz, limited only by the pulse repetition frequency of the system (Supplementary Note 3 and Supplementary Fig. 7). As a preliminary work, chirp excitation was tested with promising results as shown in Supplementary Fig. 8.

## Methods

**Ultrasonic antenna fabrication.** A 1 mm thick bulk PZT plate was purchased from American Piezo (APC - 855). To achieve the desired dimensions, the PZT plate was diced using a wafer dicer (DAD3221, DISCO). The 200 µm thick flexible conductive substrate, ML-104, was purchased from LPKF. The substrate was patterned and a 2.6×2.6 mm$^2$ area was cut for air-backing design using an R4 laser machining device (LPKF Laser & Electronics). The top and bottom electrodes of the diced PZT crystal were soldered manually to the copper pads on the patterned ML-104 substrate using a low-temperature solder (TS391LT, Chip Quik). The air cavity at the back of the PZT crystal was sealed by placing a Kapton tape (3M) on the back of the flexible substrate. After electrical connections between the US antenna and the capacitive sensor were established, the device was coated with Parylene-C (PPS Labcoater Series 100) for an hour, creating a 900 nm thick conformal encapsulation for electrical and chemical isolation.

**Device characterization.** The electrical and acoustic characterizations were performed in a distilled water environment. An impedance analyzer (MFIA, Zurich Instruments) was used to acquire the frequency-dependent complex electrical impedance of the US antenna. A 0.5 mm needle hydrophone (Precision Acoustics) was used for the acoustic characterizations. An arbitrary waveform generator (33521B Waveform Generator, Keysight) was used to excite the US antenna, while a motorized ultrasonic measurement system (UMS Research, Precision Acoustics) controlled the position of the needle hydrophone connected to a digital oscilloscope (DSOX3024G, Keysight). The acoustic pressure field scan was carried out on 25×25 grid points with 0.4 mm increments. Fig. 2h shows the interpolated version of the 25×25 data.

**Communication method characterizations.** The characterization tests of the passive ultrasonic communication method were conducted in a distilled water environment using a commercial immersion transducer (V323-SU, Olympus) as the interrogator transducer. The load capacitance on the US antenna was varied by connecting the antenna to a variable capacitor (PPZN60100, Passive Plus). The capacitance of the variable capacitor was measured by an impedance analyzer (MFIA, Zurich Instruments) after each adjustment. The excitation and signal processing operations of the communication sweeps were done by a lock-in amplifier equipped with arbitrary waveform generation (AWG) capability (UHFLI, Zurich Instruments). A duplexer box (RDX-6, RITEC Inc.) was used to enable the pulse-echo operations. The post-processing was done using Python scripts.

**Communication method scheme.** The signal output channel of the lock-in amplifier (LIA) is connected to an RF power amplifier (F30PV, Pendulum) whose output port is connected to the duplexer box via a BNC cable. The receiver signal port of the duplexer is directly connected to the LIA input signal channel.

The pulse-echo scheme is realized using the lock-in amplifier, which facilitates an arbitrary waveform generator (AWG) module to create 12 μs long modulated electrical pulses to drive the interrogator transducer. An internal oscillator determines the modulation frequency. The device's echo is demodulated using this oscillator as well. This scheme allows sweeping the oscillator frequency while keeping the measurement and pulse modulation in sync. The frequency sweep is performed in the vicinity of the anticipated anti-resonance frequency. At each frequency point, the signal is demodulated using a 4th order low-pass filter with 200 kHz cutoff frequency. The resulting signal is the envelope of the modulated signal corresponding to the backscattered wave as shown in Fig. 3b. The control software calculates the average amplitude and plots it against the frequency. The sweep results is continuously fed into a Python script, which applies a 1D Gaussian filter to smooth the spectrum before outputting the frequency of the anti-resonance valley (Supplementary Fig. 9).

**In-vitro demonstration of the communication method.** The in-vitro experiment setup was constructed using a 15×15×15 $cm^3$ polycarbonate box and a 3D-printed top cover, which holds the interrogator transducer and facilitates the air hose connections. Rubber o-rings were used to seal the box and prevent air leakages. A total of three air hose connections were designed, one for the air inlet, one for the air outlet, and one for the reference air pressure sensor readings.

The commercial pressure sensor (FlexiForce A201, Tekscan) was integrated to the fabricated US antenna and the integrated device was placed inside the box such that the US antenna is directly under the interrogator transducer. The sensing area of the commercial pressure sensor was fixed on the box wall. The reference pressure readings were taken by a commercial air pressure sensor (NUL210, NeuLog).

**Acknowledgements**

U.C.Y. is supported by TUBITAK through the 2210/A programme. A.T.T is supported by Scientific and Technological Research Council of Turkey (TUBITAK) through the 2232 (grant no. 118C295) programme. L.B. acknowledges TUBITAK 2232 (grant no. 118C295) and the European Research Council (grant no. 101043119). We acknowledge Koç University Nanofabrication and Nanocharacterization Center (n2STAR) for access to the infrastructure.

**Author Contributions**

U.C.Y. and L.B. conceived the research idea. U.C.Y. designed, fabricated, and characterized the ultrasonic antenna, conducted experiments, and analyzed the data. A.T.T. contributed to the experiment design and technical discussions. K.E. and M.A. contributed to signal processing. L.B. acquired funding and directed the research activities. All authors contributed to the manuscript writing.

**Competing Interests**

The authors declare no competing interests.

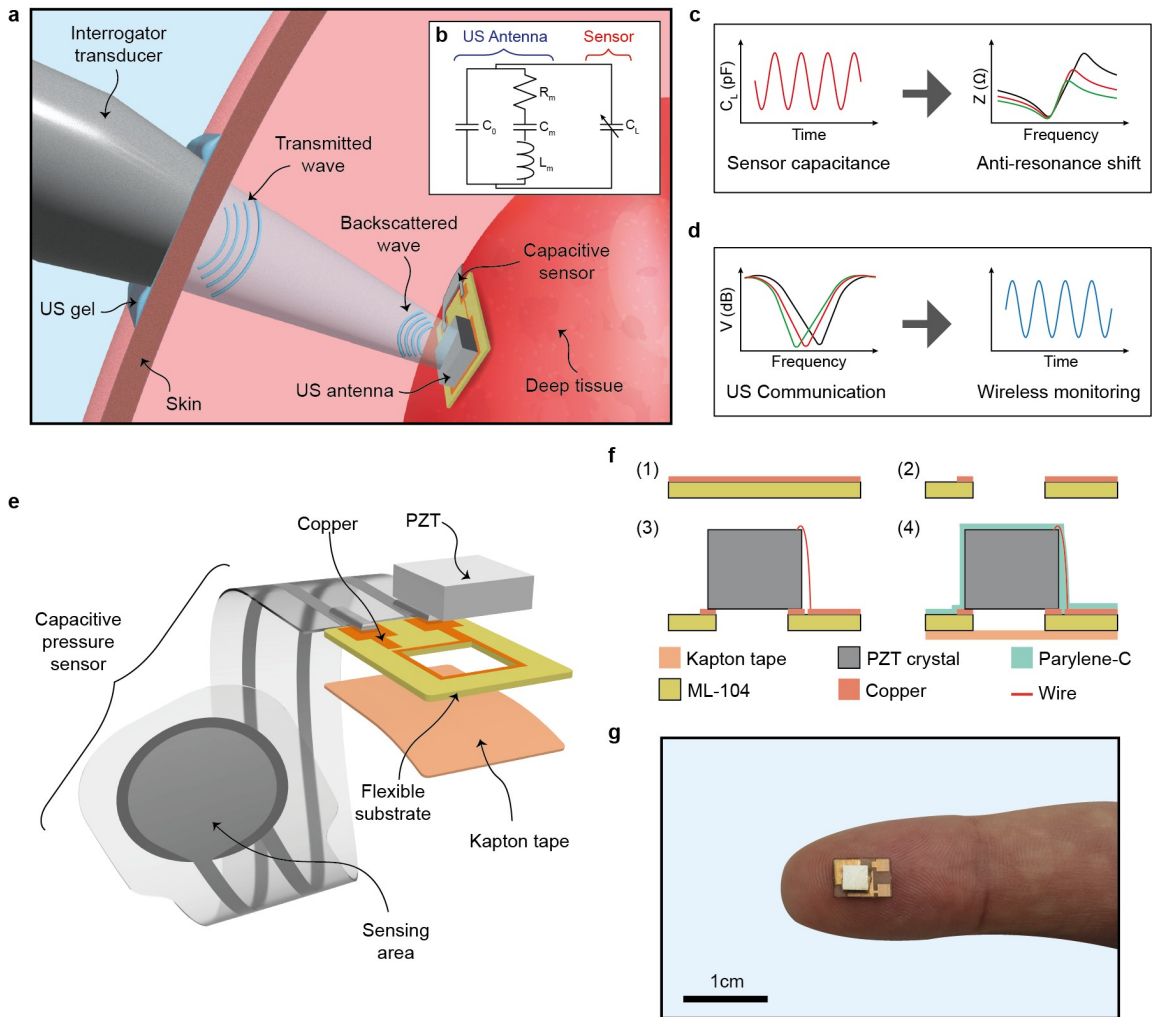

**Fig. 1 | Device design and working principle. a** Schematic of the passive ultrasonic communication method. The ultrasonic antenna is placed on a deep tissue and the wireless communication link is established between the device and the external interrogator transducer. **b** The simple equivalent circuit model of the implanted device. The ultrasonic antenna and capacitive sensor are represented by a van-Dyke Butterworth circuit and a capacitor, respectively. **c** Deep tissue measurements change the sensor capacitance, which directly affects the anti-resonance frequency of the device. **d** Anti-resonance frequency shifts are detected wirelessly using the passive ultrasonic communication method. The shifts occurring in the frequency spectrum are translated into wireless sensor readings. **e** Exploded view of the device, showing the patterned flexible conductive substrate, piezoceramic antenna, and air-cavity encapsulation. **f** The fabrication flow starts with an unpatterned conductive substrate (1), which is cut and patterned using laser micromachining (2). The PZT crystal is then soldered on the copper traces (3). As a last step, the air cavity is sealed using Kapton tape, and the device is encapsulated using Parylene-C (4). **g** Optical image of the PUC device on a finger for size comparison before the top electrode is soldered.

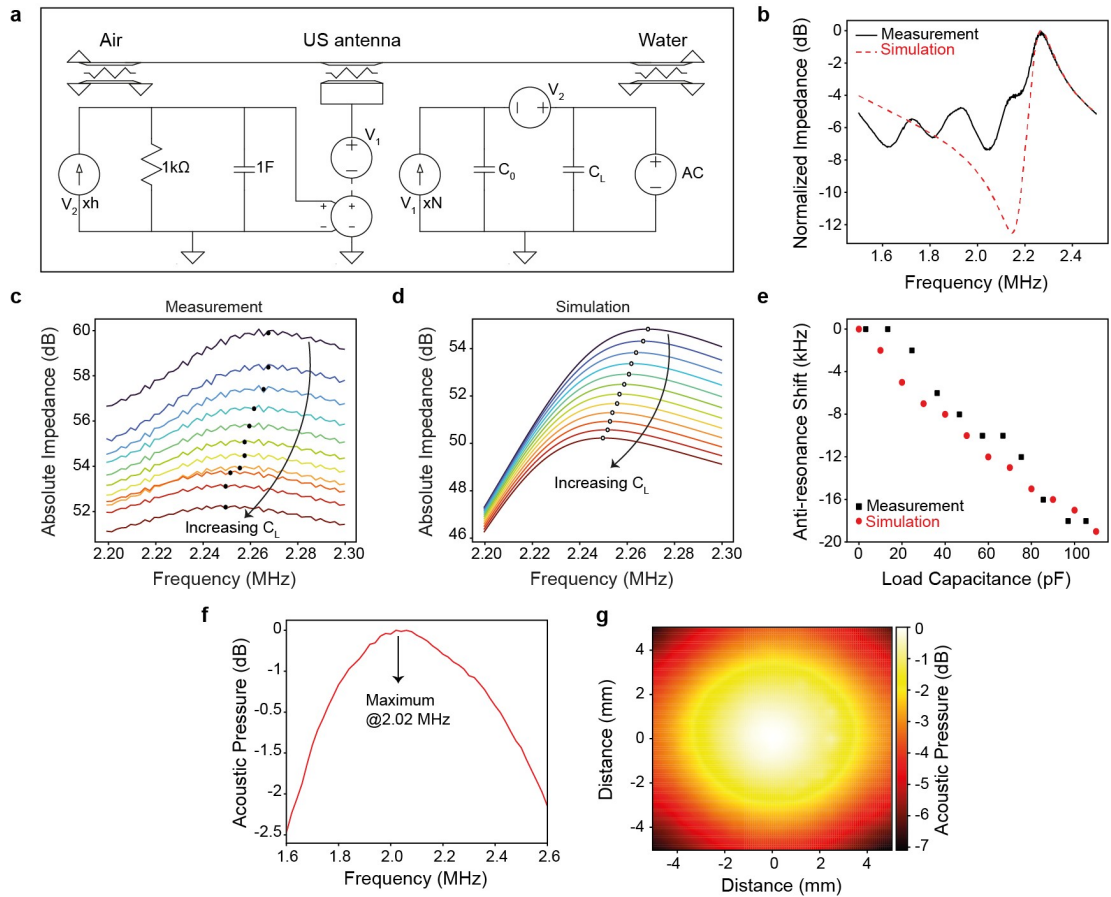

**Fig. 2 | Electrical and mechanical device characterization of the device. a** Schematic of the SPICE equivalent circuit model of the ultrasonic antenna in water. The Leach model is used for its SPICE-friendly nature. The air-backing medium, PZT crystal, and the acoustic medium are modeled as lossy transmission lines. **b** Measured and simulated absolute electrical impedance of the US antenna. The resonance characteristics of the fabricated device do not match fully with the simulation due to unconventional PZT dimensions. **c,d** Electrical impedance curves of the measured (**c**) and simulated (**d**) US antenna under various capacitive loads. The peaks in the impedance curves are anti-resonance frequencies. **e** Anti-resonance frequency shifts for varying capacitive loads. The measurement results show good agreement with the simulation. **f** Acoustic frequency sweep test results of the US antenna, showing peak acoustic pressure level at 2.02 MHz. **g** Acoustic pressure field of the US antenna is measured by a needle hydrophone at a depth of 3 cm.

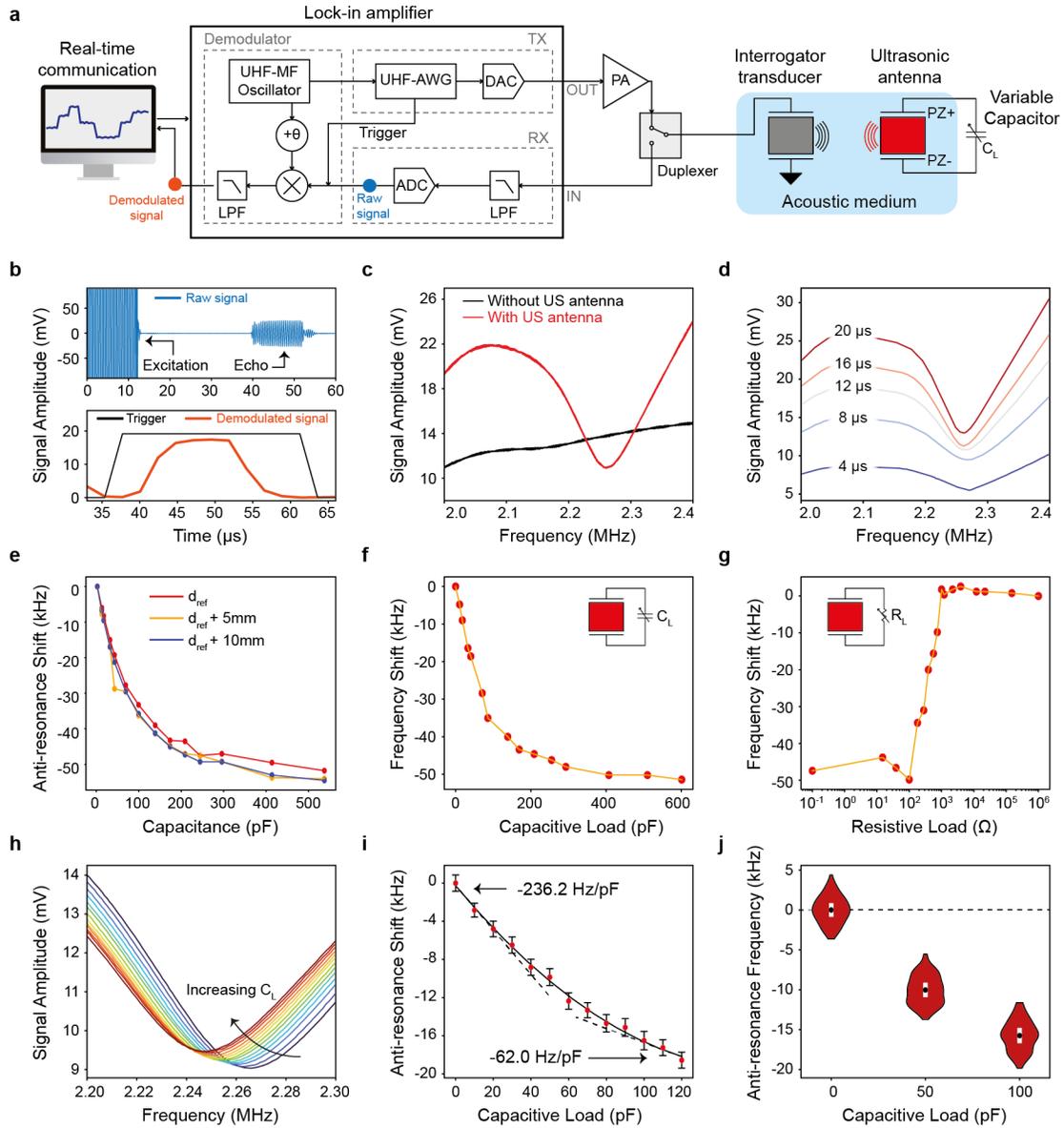

**Fig. 3 | Communication link characterization. a** Schematic of the wireless communication hardware block diagram. The pulse-echo operation is enabled by using a duplexer box and the signal processing is handled by a lock-in amplifier. **b** The raw and demodulated time-domain waveforms of a pulse-echo operation. The lock-in amplifier calculates the signal level of the echo by averaging the data points within the demodulator trigger interval. The system repeats this process for every frequency within the specified interval and forms a frequency amplitude spectrum. **c** The acquired frequency spectrum when the US antenna is in front of the interrogator transducer is compared with the spectrum acquired without the US antenna (n=10 measurements). **d** The effect of the excitation signal length on the communication sweep profile. **e** Anti-resonance shifts acquired for various antenna-transducer distances. **f-g** The frequency shifts when capacitive (f) and resistive (g) loads are connected to the US antenna (A 10x10 mm$^2$ US antenna was used for e-g, n=3 measurements). **h** Frequency spectrums acquired by the passive ultrasonic communication method for varying capacitive load connected to the antenna. **i** The frequency of the valley (anti-resonance) shifts as the load capacitance increases (n=50

measurements). **j** Violin plot of the wireless communication method, showing the precision of the communication link. The mean and standard deviation of the measurements are shown with white points and black bars, respectively (n=100 measurements).

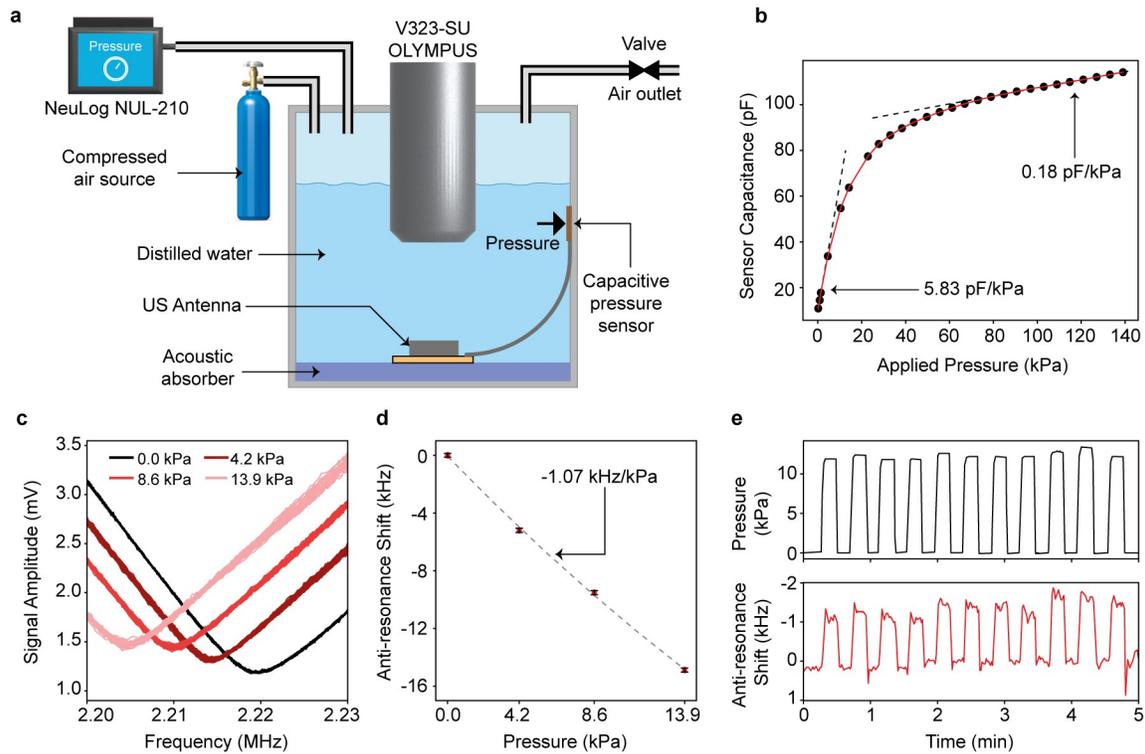

**Fig. 4 | In-vitro demonstration of the PUC system. a** The schematic of the in vitro experiment setup showing the compressed air connections, the interrogator transducer and the US antenna, the integrated capacitive pressure sensor, and the commercial air pressure sensor. **b** The capacitance plot of the integrated capacitive pressure sensor under varying pressure. **c** Communication sweep curves for four pressure levels before applying post-processing (n=40 measurements). **d** The valley frequency shifts yield a sensitivity of approximately -1.07 kHz/kPa. **e** Cyclic performance of the communication method. The pressure in the tank was varied manually by controlling the compressed air valve. The communication sweeps were recorded in real-time with a sample rate of 1 Hz. The anti-resonance shift in the cyclic test was less than anticipated due to transverse and angular misalignment effects between the transducer and the US antenna.